\documentclass{sf2a-conf2011}
\usepackage{graphicx}
\usepackage{hyperref}
\usepackage[]{natbib}  
\usepackage[cyr]{aeguill}
\usepackage{epstopdf}

\def\BibTeX{{\rm B\kern-.05em{\sc i\kern-.025em b}\kern-.08em
    T\kern-.1667em\lower.7ex\hbox{E}\kern-.125emX}}
\bibpunct{(}{)}{;}{a}{}{,}  

\begin{document}

\TitreGlobal{SF2A 2011}


\title{The formation of large galactic disks: revival or survival?}

\runningtitle{Formation of large galactic disks}

\author{F. Hammer}\address{GEPI, Observatoire de Paris, CNRS, 5 Place Jules Janssen, 92195 Meudon, France}


\author{Puech, M. $^1$}
\author{Flores, H. $^1$}

\author{Athanassoula, E.}\address{LAM, CNRS/UMR6110 and Universit\'e de 
Provence,
  38 rue Fr\'ed\'eric Joliot-Curie, 13388 Marseille C\'edex 13, France} 

\author{Yang, Y. B. $^1$}
\author{Wang, J. L. $^1$}
\author{Rodrigues, M. $^1$}
\author{Fouquet, S. $^1$}

\setcounter{page}{237}

\index{Hammer, F.}


\maketitle


\begin{abstract}
Using the deepest and the most complete set of observations of distant
galaxies, we investigate how extended disks could have
formed. Observations include spatially-resolved kinematics, detailed
morphologies and photometry from UV to mid-IR. Six billion years ago,
half of the present-day spiral progenitors had anomalous kinematics
and morphologies, as well as  relatively high gas fractions. We argue
that gas-rich major mergers, i.e., fusions between gas-rich disk
galaxies of similar mass, can be the likeliest driver for such strong
peculiarities. This suggests a new channel of disk formation,
e.g. many disks could be reformed after gas-rich mergers. This is
found to be in perfect agreement with predictions from the
state-of-the-art $\Lambda$CDM semi-empirical models: due to our
sensitivity in detecting mergers at all phases, from pairs to relaxed
post-mergers, we find a more accurate merger rate. The scenario can be finally  confronted to properties of nearby galaxies, including M31 and galaxies showing ultra-faint, gigantic structures in their haloes. 
\end{abstract}



\section{Introduction}
Seventy two percent of local galaxies with $M_{stellar}>$ 2$\times$
$10^{10}$ $M_{\odot}$ are disk-dominated. Thin disks are fragile to
collisions with other galaxies that can easily destroy them
\citep{Toth92}. $\Lambda$CDM predicts a high level of merger activity
on all scales and this makes it difficult for the corresponding simulations to reproduce such a large number of large disks with small bulge fraction. This is illustrated by the tidal torque theory \citep{Peebles76,White}, which assumes that the angular momentum of disk galaxies had been acquired by early
interactions: related simulations provide too small disks with too small angular momentum, compared to observations.

What do we learn from observations? The situation seems somewhat confused with discordant results on the real impact of mergers, either minor or major. Thus, the role of mergers in the evolution of disk galaxies remains uncertain. Discordant results can be attributed to the following reasons:
\begin{itemize}
\item differences between galaxy sample selection, for example various stellar/baryonic mass ranges;
\item different methodologies to characterise a merger (pair technique, automatic classification methods such as concentration-asymmetry and GINI-M20, decision tree methods, etc.);
\item different methodologies to classify "normal" galaxies, especially disk galaxies: either automatic classification or systematic comparison to local templates and use of spatially-resolved kinematics to verify the presence of rotation;
\item the depth and spatial resolution of the images to which the above methodologies have been applied.
\end{itemize}

In this paper, we present the results of the IMAGES survey that encompasses the deepest and the most complete set of measurements of galaxies at z=0.4-0.8.  The explicit goal of IMAGES is to gather enough constraints z=0.4-0.8 galaxies to directly link them to their descendants, the local galaxies. Its selection is limited by an absolute J-band magnitude ($M_{J}(AB)<$ -20.3), a quantity relatively well linked to the stellar mass \citep[][hereafter IMAGES-I]{Yang08}, leading to a complete sample of 63 galaxies with $M_{stellar}>$ 1.5 $10^{10}$ $M_{\odot}$, and with an average value similar to the Milky Way mass. The set of measurements includes:
\begin{itemize}
\item  ACS imagery from GOODS (3 orbits in b, v, i and z) to recover colour-morphology comparable to the depth and resolution ($<$400 pc) of the SDSS \citep{Delgado09};
\item spatially-resolved kinematics from FLAMES/GIRAFFE (from 8 to 24hrs integration) to sample gas motions at $\sim$ 7 kpc resolution scale (IMAGES I); 
\item deep VLT/FORS2 observations (3hrs with two grisms at R=1500) to recover the gas metal abundances \citep[][hereafter IMAGES-IV]{Rodrigues08};
\item Spitzer 24$\mu$m observations of the GOODS field to estimate the extinction-corrected star formation rates as well as Spitzer IRAC and GALEX deep observations of the field to provide photometric points to constraint the spectral energy distribution.
\end{itemize}

Taken together, these measurements ensure that the IMAGES sample is collecting an unprecedented amount of data with depth and resolution comparable to what is currently obtained for local galaxies. For example IMAGES is hardly affected by cosmological dimming, because 3 HST/ACS orbits ensure the detection of the optical disk of the Milky Way after being redshifted to z $\sim$ 0.5. In section 2 we present the morphological and kinematical properties of distant galaxies, and propose a scenario to relate them to their present-day mass analogues. In section 3 we discuss these results in the context of the $\Lambda$CDM model. In section 4 we verify whether this link is robust when compared to the detailed observations of nearby spirals and their haloes.

\section{What is the past history of giant spiral galaxies?}
Progenitors of present-day giant spirals are similar to galaxies having emitted their light $\sim$ 6 Gyr ago, according to the Cosmological Principle. The IMAGES sample is then unique to sample these progenitors: Fig. 1 presents the results of a morphological analysis of 116 SDSS galaxies (top) and of 143 distant galaxies including those from IMAGES (bottom) for which depth, spatial resolution and selection are strictly equivalent \citep{Delgado09}. Methodology for classifying the morphologies follows a semiautomatic decision tree, which uses as templates the well known morphologies of local galaxies that populate the Hubble sequence, including the color of their sub-components \citep[][see their Fig. 4]{Delgado09}. Such a conservative method is the only way for a robust morphological classification, and indeed, the \cite{Delgado09} results are similar to those of experts in the field \citep[e.g.,][]{vBergh2002}. The second step in classifying the nature of distant galaxies is to compare the morphological classification to the spatially-resolved kinematics. The latter provides a kinematical classification of velocity fields ranging from rotation, perturbed rotation or complex kinematics \citep{Flores06,Yang08}. \cite{Neichel08} robustly established that peculiar morphologies coincide well with anomalous kinematics and vice versa: 95\% and 86\% of galaxies with complex kinematics  and perturbed rotations have peculiar morphologies, respectively. On the other hand 80\% of galaxies with robust rotation show spiral morphologies.

\cite{Neichel08} also verified whether such a situation is preserved when using automatic classification methods such as concentration-asymmetry and GINI-M20. The answer is negative, and these methods overestimate the number of spirals by a factor of two, a problem already identified by \cite{Conselice2005}. Such methods are interesting because they can be applied to a much larger number of galaxies than the 143 galaxies studied by \cite{Delgado09}. However their limitations in distinguishing peculiar from spiral morphologies lead to far larger uncertainties than the Poisson statistical noise in \cite{Delgado09}. 

\begin{figure}[ht!]
 \centering
 \includegraphics[width=0.7\textwidth,clip]{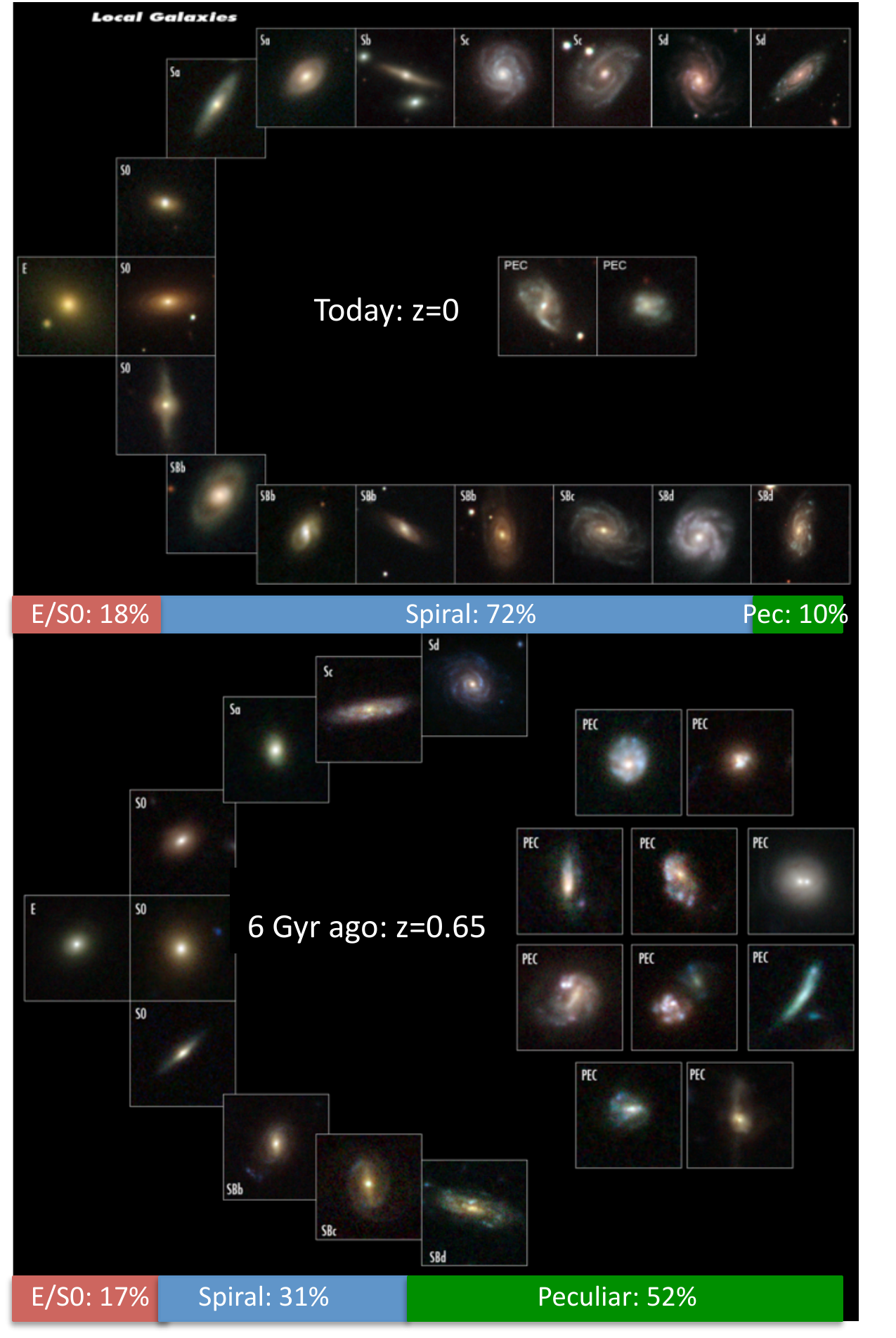}      
  \caption{An adaptation of the \cite{Delgado09} Figure 5: Present-day Hubble sequence derived from the local sample (top) and past Hubble sequence derived from the distant sample (bottom). Each stamp represents approximately 5\% of the galaxy population. Galaxy fractions are given in percentage.}
  \label{author1:fig1}
\end{figure}

Fig. 1 presents the global evolution of the Hubble sequence during the
past 6 Gyr. The link between the two Hubble sequences (past and present-day) is marginally affected by very recent mergers (number density
evolution) or by stellar population evolution (luminosity or stellar
mass evolution).  While the former is limited by the expected decrease
of mergers at recent epochs, the latter is precisely compensated by
the evolution of the $M_{stellar}$/$L_{K}$ ratio \citep[][see their
  sect. 5.3]{Delgado09}. As a results the fraction of E/S0 has not
evolved, while half of the spirals were not in place 6 Gyr ago, or in
other words, half of the spiral progenitors have either peculiar
morphology and/or anomalous kinematics.

The remarkable agreement between morphological and kinematical
classifications implies that dynamical perturbations of the gaseous
component at large scales are linked to peculiar morphological
distribution of the stars. This indicates a common process at all
scales for gas and stars in these galaxies. Which physical processes
may be responsible of this morpho-kinematic behaviour? Most anomalous
galaxies reveal peculiar large-scale gas motions that cannot be caused
by minor mergers: although they can affect locally the dispersion map they do not affect the large scale
rotational field over several tens of kpc \citep{Puech07}. Internal fragmentation is
limited because less than 20\% of the sample show clumpy morphologies
according to \cite{Puech10} while associated cold gas accretion tends
to vanish in massive halos at z$<$1, with $<$1.5 $M_{\odot}$/yr at
z$\sim$ 0.6 \citep{Keres09, Brooks09}. Finally perturbations from
secular and internal processes (e.g. bars or spirals) are too small to be detected by the "large-scale" spatially resolved spectroscopy of IMAGES. Major mergers appear to be the most likely mechanism to explain the above properties of anomalous galaxies and indeed it is the only way to explain the strong redshift evolutions of the scatter of the Tully-Fisher relation \citep{Puech08,Puech10b} and of the scatter of the luminosity-metallicity relation \citep{Hammer01,Liang06,Rodrigues08}.

This has led our team to test and then successfully model five of the
IMAGES galaxies as consequences of major mergers
\citep{Peirani09,Yang09,Hammer09b,Puech09,Fuentes10} using
hydrodynamical simulations (GADGET2 and ZENO). However the amount of
data to be reproduced per galaxy is simply enormous, leading to 21
observational constraints to be compared to 16 free model parameters
in the specific case of \cite{Yang09}. We have then limited our
subsequent analysis to the 33 galaxies belonging to the CDFS for
reasons of data homogeneity. A comparison of their morpho-kinematics
properties  to those from a grid of simple major merger models based
on \cite{Barnes02}, provided convincing matches in about two-thirds of
the cases. This implies that a third of z=0.4-0.75 spiral galaxies are
or have been potentially involved in a major merger. Since major mergers can easily destroy thin rotating disks, this creates an apparent tension between the large fraction of present-day disks and their survival within the $\Lambda$CDM \citep[e.g.,][]{Stewart09}. On the other hand this appears consistent with expectations from \cite{Maller02} i.e., that: "the orbital angular momentum from major mergers may solve the spin catastrophe". 

\section{Are disks surviving or reviving after a merger?}

\cite{Barnes02} described the re-formation of disks after major
mergers, assuming a Milky Way gas fraction (12\%) in the
progenitors. With larger gas fraction the rebuilt disk can be more prominent and in case of extremely high gas fraction, could dominate the galaxy \citep{Brook04,Springel05,Robertson06}. After a gas-rich merger a prominent gaseous disk can form, which could be the progenitor of some present-day disks. These models could appear to be of relatively limited significance given the very large assumed gas fractions (up to 90\%). In fact observations of distant galaxies indicate gas fractions that may exceed 50\% at z$\sim$ 1.5-2 \citep{Daddi10, Erb06}, and it is unclear whether or not higher gas fractions may be common at those redshifts. It could be possible to circumvent this difficulty, perhaps through tuning some physical ingredients in the models, e.g. a feedback more efficient within the central region \citep{Governato09}, or a star formation less efficient at earlier epochs \citep{Hammer10}. Such methods are not necessarily wrong, but their additional value is limited since they have been designed intentionally to preserve the gas before the merger or to remove the gas from the central regions to redistribute it to the newly formed disk. 

Important progress is expected on both observational and theoretical sides: a confirmation/infirmation of the IMAGES result has to be done by an independent team, although it needs to avoid automatic procedures that often degrade the significance of astrophysical data. With this,  the so numerous unstable, anomalous progenitors of present-day spirals, with behaviour so similar to major mergers will be the major constraint for disk galaxy formation theories. Why IMAGES is finding so many galaxies (a third to half of the spiral progenitors) that can be attributed to a major merger phase? In fact the morpho-kinematic technique used in IMAGES is found to be sensitive to all merger phases, from pairs to post-merger relaxation.  \cite{Puech11} has compared the merger rate associated with these different phases, and found a perfect match with predictions by state-of-the-art $\Lambda$CDM semi-empirical models \citep{Hopkins10} with no particular fine-tuning. Thus, both theory and observations predict an important impact of major mergers for progenitors of present-day spiral galaxies: the Hubble sequence made of elliptical and spiral galaxies could be just a vestige of merger events \citep{Hammer09b}.

\cite{Athanassoula10} described the different physical processes
leading to the formation of elliptical and spiral galaxies. Since the "merger hypothesis" by 
\cite{Toomre72}, it is often accepted that elliptical galaxies may be
the product of a major merger between two gas-poor spiral galaxies. It appears more and more plausible that some spiral galaxies could also result from a gas-rich merger of two smaller spiral galaxies. An increasing number of cosmological simulations lead to the formation of late-type disk galaxies after major mergers \citep{Font11, Brook11}. There are still some important questions on precisely how galaxies dominated by thin disks can be generated by such violent processes. We also need to examine whether this scenario can be reconciled with observations of large disks in present-day spiral galaxies.

\section{Can a rich merger history be reconciled with observations of nearby spirals?}
Having a tumultuous merger history 6 Gyr ago should have left some imprints in many present-day spiral galaxies. Let us consider our nearest neighbour, M31. Quoting \citet{vandenBergh05}: ``Both the high metallicity of the M31 halo, and the $r^{1/4}$ luminosity profile of the Andromeda galaxy, suggest that this object might have formed from the early merger and subsequent violent relaxation, of two (or more) relatively massive metal-rich ancestral objects.'' In fact the considerable amount of streams in the M31 haunted halo could be the result of a major merger instead of a considerable number of minor mergers \citep{Hammer10}. This alternative provides a robust explanation of the Giant Stream discovered by \cite{Ibata01}: it could be made of stars returning from a tidal tail that contains material previously stripped from the lowest mass encounter prior to the fusion.  In fact stars in the Giant Stream \citep{brown07} have ages older than 5.5 Gyr, which is difficult to reconcile with a recent collision that is expected in a case of a minor merger \citep[e.g.,][]{Font08}. This constraint has let \cite{Hammer10} to reproduce the M31 substructures (disk, bulge \& thick disk) as well as the Giant Stream after a 3:1 gas-rich merger for which the interaction and fusion may have occurred 8.75$\pm$0.35 and 5.5 $\pm$0.5 Gyr ago, respectively. Besides this, the Milky Way may have had an exceptionally quiet merger history \citep[e.g.,][]{Hammer07}.

Further away from the Milky Way, \cite{Martinez10} conducted a pilot survey of isolated spiral galaxies in the Local Volume up to a low surface
brightness sensitivity of $\sim 28.5$ mag/arcsec$^2$ in the $V$ band. They found that many of these galaxies have loops or streams of various shapes. These observations are currently considered as evidencing the presence of minor mergers in spiral galaxies. For example, NGC5907 is showing the most spectacular loops thar have been modelled by a very minor merger (mass ratio is 4000:1) by \cite{Martinez08}. Instead of that, \cite{Wang11} recently succeed to model the NGC5907 galaxy and their associated loops by assuming a 3:1 gas-rich major merger during the past 8-9 Gyr, for which the loops are caused by returning stars from tidal tails.

There is still a considerable work to do to establish firmly which
process is responsible for the tumultuous history of nearby spirals
that is imprinted into their haloes. In most cases \citep{Martinez10},
there is no hint of the residual of the satellite core that is
responsible of the faint structures discovered in the nearby spiral
haloes. If confirmed, this may be problematic for the minor merger
scenario. On the other hand, the major merger alternative still faces
the problem of reconstructing thin disks that are consistent with the
observed ones. However, numerical simulations are rapidly progressing, and AREPO-like simulations \citep{Keres11} provide much higher resulting angular momentum when compared to GADGET, and thus thin disks that could resemble much more to the observed ones.  Another important advance is provided by Spitzer observations of edge-on spirals \citep{Comeron11}, indicating more massive thick disks than previously reported, these structures being naturally expected in the case of major mergers.

\section{Conclusion}
The first proposition that major mergers could be responsible of the
re-formation of $\sim$ 70\% of present-day galactic disks
\citep{Hammer05}, was at that time only based on the coeval evolution
of morphologies, star formation density and merger rate. Subsequent
morpho-kinematic analyses are providing a much more robust
confirmation and accuracy to this scenario. It now receives much more
attention from both the theoretical side -- with an impressive number
of articles aiming at reforming spiral galaxies after a collision --
and from the observational side -- with a large number of papers discussing the influence of mergers in galaxy formation. 

Here we plead for the use of a complete set of (observationally), well-determined parameters to characterise distant galaxies. Galaxies are made of hundred billions of stars and distant galaxies contain an equivalent amount of mass of gas. As such they are complex objects and, to be relevant, analyses should include detailed characterisations of their morphologies, kinematics, star formation and gas and stellar masses. Very large surveys are very powerful in gathering huge number of galaxy spectra, although they often lead to oversimplifications related to automatic procedures in characterising galaxies. 

Having characterised distant galaxies with unprecedented details through the IMAGES project, this supports that a third to a half of spiral progenitors were in a merger phase at z=[0.4-0.75]. This can potentially reconcile the  $\Lambda$CDM scenario, predicting a large fraction of mergers, with the very large fraction of large disks in present-day galaxies with masses similar to that of the Milky Way. Consequences of a disk reformation after a merger episode could have important impacts in modern cosmology.

\begin{acknowledgements}
F.H. thanks Collen Sharkey and his team at the Hubble European Information Centre for their remarkable work to disseminate science and for the reprocessing of the Fig. 5 of Delgado et al. (2010). We thank Yanchun Liang and Benoit Neichel  who have noticeably contributed to the success of the IMAGES project.
\end{acknowledgements}


\bibliographystyle{aa}  
\bibliography{sf2a-template} 

\end{document}